\documentclass[useAMS,usenatbib,epsfig]{mn2e}
\newif\ifAMStwofonts
\AMStwofontstrue
\usepackage{amsmath,amsfonts,epsfig,natbib}

\def\reff@jnl#1{{\rm#1\/}}
 
\def\aj{\reff@jnl{AJ}}                  
\def\araa{\reff@jnl{ARA\&A}}            
\def\apj{\reff@jnl{ApJ}}                        
\def\apjl{\reff@jnl{ApJ}}               
\def\apjs{\reff@jnl{ApJS}}              
\def\ao{\reff@jnl{Appl.Optics}}         
\def\apss{\reff@jnl{Ap\&SS}}            
\def\aap{\reff@jnl{A\&A}}               
\def\aapr{\reff@jnl{A\&A~Rev.}}         
\def\aaps{\reff@jnl{A\&AS}}             
\def\azh{\reff@jnl{AZh}}                        
\def\baas{\reff@jnl{BAAS}}              
\def\jrasc{\reff@jnl{JRASC}}            
\def\memras{\reff@jnl{MmRAS}}           
\def\mnras{\reff@jnl{MNRAS}}            
\def\pra{\reff@jnl{Phys.Rev.A}}         
\def\prb{\reff@jnl{Phys.Rev.B}}         
\def\prc{\reff@jnl{Phys.Rev.C}}         
\def\prd{\reff@jnl{Phys.Rev.D}}         
\def\prl{\reff@jnl{Phys.Rev.Lett}}      
\def\pasp{\reff@jnl{PASP}}              
\def\pasj{\reff@jnl{PASJ}}              
\def\qjras{\reff@jnl{QJRAS}}            
\def\skytel{\reff@jnl{S\&T}}            
\def\solphys{\reff@jnl{Solar~Phys.}}    
\def\sovast{\reff@jnl{Soviet~Ast.}}     
\def\ssr{\reff@jnl{Space~Sci.Rev.}}     
\def\zap{\reff@jnl{ZAp}}                        
\def\nat{\reff@jnl{Nature}}             

\title[VSA first results I -- methods]{First results from the Very Small Array -- I.
Observational methods}

\author[R A Watson et al] {Robert A. Watson$^{1,\dagger}$,
 Pedro Carreira$^1$, Kieran Cleary$^1$, Rod D. Davies$^1$,\newauthor Richard J.
 Davis$^1$, Clive Dickinson$^1$, Keith Grainge$^2$,
 Carlos M. Guti{\'e}rrez$^3$,  \newauthor
 Michael P. Hobson$^2$, 
 Michael E. Jones$^2$, R\"udiger Kneissl$^2$,
 Anthony Lasenby$^2$, \newauthor 
 Klaus Maisinger$^2$,
 Guy G. Pooley$^2$, Rafael Rebolo$^{3,4}$, Jos\'e Alberto
 Rubi\~no-Martin$^3$, \newauthor
 Ben Rusholme$^{2,\star}$, Richard
 D.E. Saunders$^2$, Richard Savage$^2$, Paul F. Scott$^2$,\newauthor 
 An\v ze Slosar$^2$,
 Pedro J. Sosa Molina$^3$, Angela C. Taylor$^2$,  David
 Titterington$^2$,  \newauthor
 Elizabeth Waldram$^2$ and Althea Wilkinson$^1$
\\
  $^1$Jodrell Bank Observatory, University of Manchester, UK.\\
  $^2$Astrophysics Group, Cavendish Laboratory, University of
  Cambridge, UK. \\ $^3$Instituto de Astrof{\'i}sica de Canarias, 38200 La
  Laguna, Tenerife, Spain.\\ 
  $^4$Consejo Superior de Investigaciones Cient{\'{\i}}ficas, Spain \\
  $^{\dagger}$Present address: Instituto de Astrof{\'{\i}}sica de Canarias.\\
  $^{\star}$Present address:
  Stanford University, Palo Alto, CA, USA.\\
}
\date{Accepted Received In original form}
\pagerange{\pageref{firstpage}--\pageref{lastpage}}
\pubyear{2001}

\def\LaTeX{L\kern-.36em\raise.3ex\hbox{a}\kern-.15em
    T\kern-.1667em\lower.7ex\hbox{E}\kern-.125emX}

\begin{document}

\label{firstpage}

 \maketitle

\begin{abstract}

The Very Small Array (VSA) is a synthesis telescope designed to
image faint structures in the cosmic microwave background on degree
and sub-degree angular scales. 
The VSA has key differences from other CMB interferometers with the
result that different systematic errors are expected. 
We have tested the operation of the VSA
with a variety of blank-field and calibrator observations and
cross-checked its calibration scale against independent
measurements. We find that systematic effects can be suppressed below
the thermal noise level in long observations; the overall calibration
accuracy 
of the flux density scale 
is 3.5 percent and is limited by the external absolute calibration scale. 

\end{abstract}

\begin{keywords}
 cosmology:observations -- cosmic microwave background
\end{keywords}

\section{INTRODUCTION}

Sensitive measurements of the cosmic microwave background (CMB) radiation are
increasingly important in cosmology and astrophysics. Observations of both the
primary imprints on the CMB originating at the surface of last scattering and
secondary imprints such as the Sunyaev-Zel'dovich effect contain a wealth of
information, if they can be made with sufficient sensitivity and freedom from
systematic errors. We have constructed a telescope, the Very Small Array
(VSA), for measuring intensity structures in the CMB on degree and sub-degree
angular scales. The telescope itself is described in detail in
Rusholme et al. (in prep.). The present paper discusses the
observational methods 
underlying VSA  
observations, and the procedures used to eliminate systematic errors
from the data (such as
contaminating emission from the Sun). In three accompanying
papers we then present results from the first year's observations with the
VSA. Paper II \citep{VSApaperII} describes the first season's observations and
issues of foreground removal; Paper III \citep{VSApaperIII} presents the power
spectrum results and analysis, while Paper IV \citep{VSApaperIV} derives
cosmological parameters from the VSA results and other data.

Many different experimental approaches have been used to measure features in
the CMB, mostly using switched- or swept-beam systems with total power
detectors. Such systems can be made with very high sensitivity, particularly
when using broad-band bolometric detectors, and progress in the field has
mostly been due to improved techniques for suppressing unwanted 
signals such
as atmospheric emission and differential emission from the telescope
optics. An alternative approach, which we have adopted, is to use
interferometers. These provide excellent rejection of systematics, since only
signals entering both antennas with the correct path and phase modulation are
detected while signals such as ground radiation and atmospheric emission are
strongly suppressed (eg \citet{church-95}). 
For similar reasons they are  relatively insensitive to receiver gain
fluctuations.
Interferometric systems also offer the opportunity to target a
specific range of angular scales on the sky, determined by the spacing 
and the size of the
elements of the array, and are well-suited to measuring the power spectrum of
the CMB since they directly sample the Fourier modes on the sky which can then
be converted to a power spectrum. Their main disadvantages are the restricted
bandwidth compared to bolometers, due to the need for coherent receivers, and
the relative complexity and expense of the correlator. We have previously
demonstrated the success of the interferometric approach with the detection of
CMB fluctuations with the Cosmic Anisotropy Telescope (CAT) from a sea-level
site (\citet{scott-96}, \citet{baker-99}); the VSA is a direct
descendant of CAT. 

\section{THE INSTRUMENT}

In this section we briefly describe the VSA and its capabilities, and
compare it with other current CMB telescopes. 

\subsection{VSA main array}

\begin{figure}
\epsfig{file=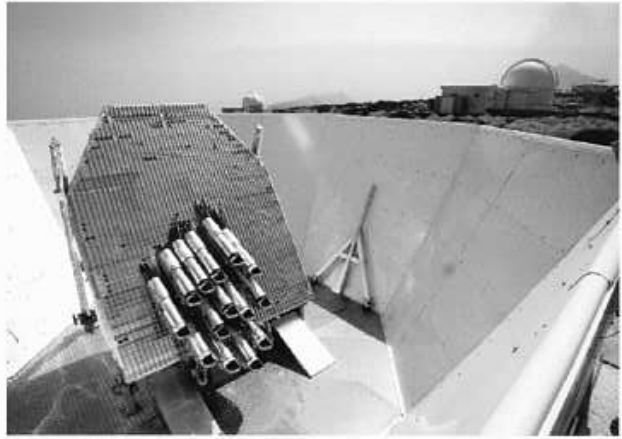,angle=270,width=8.5cm}
\caption{The VSA main array inside its ground screen. The small,
compact array antennas are fitted. \label{picture}}
\end{figure}

The VSA is a 14-element heterodyne interferometer array, tuneable between 26
and 36 GHz with a 1.5 GHz bandwidth and a system temperature of approximately
$30\,\rm{K}$, sited at 2400~m altitude at the Teide Observatory in
Tenerife. Each 
receiving element consists of a corrugated horn-reflector antenna (CHRA)
feeding a cooled HEMT amplifier, measuring one linear polarisation.
The CHRA~\citep{ghassan} consists of a
conical corrugated horn feeding a section of a paraboloidal mirror at
$90^{\circ}$ to the horn axis. This arrangement gives a compact antenna
with an unblocked aperture, and hence low sidelobes. The mirror can be rotated
about the axis of the feedhorn, allowing tracking in one dimension. The
receivers are mounted on a tilting table hinged along its northern edge,
providing tracking in a second dimension (see Figure \ref{picture}). In order
to achieve close packing of the antenna apertures, the receivers are mounted
at an angle of $35^{\circ}$ to the table; the elevation range of the table is
$0^{\circ}$--$70^{\circ}$ and thus gives a range of zenith angle of $\pm
35^{\circ}$ in 
the north-south direction, with the rotating mirrors providing approximately
$\pm 45^{\circ}$ in the other diection. The array is surrounded by an aluminium
ground screen to limit interference and groundspill; the internal sides are
sloped to direct rays from the table into the sky.

The antennas used for the observations reported here and in papers II--IV have
illuminated apertures of 143 mm diameter, giving a primary (envelope)
beam of $4.6^{\circ}$ FWHM at 34 GHz and are designed for use in a
``compact array'', for observations up to 
a maximum spherical harmonic multipole of
about $\ell = 900$. For
subsequent observations at higher $\ell$ the antennas have been modified
by increasing the length of the horn and fitting a larger mirror,
giving an aperture of 322 mm and a beam of $2.0^{\circ}$ FWHM;
observations are currently underway with this ``extended array''.

The signals are downconverted at the antennas to a first IF frequency
centred at 9~GHz, and again at a point just outside the enclosure to a
centre frequency of 1~GHz. This double downconversion allows the RF
band to be tuned by changing the first local oscillator frequency
without the need for high-Q tunable image filters. The first LO is
also phase-switched with orthogonal switching functions at each
antenna to allow subsequent phase-sensitive detection in the
correlator. The second downconversion uses LOs in quadrature to
produce real and imaginary channels for each antenna.

The second IF signals are carried on cables inside a screened
enclosure within the control building. To compensate for the changing
path to the source as the antennas track, lengths of delay line are
switched into each IF channel, eight units providing up to 1.8~m delay
in steps of 7~mm. After further gain, equalisation, and automatic gain
control, the signals are split with a cascade of power splitters and
fed into the corrrelators. Each real or imaginary correlation is done
on a separate board, where the product of the inputs is formed and
detected synchronously with the first LO switching functions. The data
are integrated in an analogue integrator for 1~s before being sampled
and read out into the control computer. The synchronous detection
circuit also includes a switch which reduces the correlator gain by a
factor of about 20, increasing the dynamic range of the correlator
for observation of very bright calibrators such as the Moon.

\subsection{Source subtraction baseline}

The most important foreground contaminant for CMB observations at
these angular scales and frequency is radio sources. The effect of
sources on the power spectrum increases as $\ell^2$, but even at the
lower $\ell$'s sampled by the compact array sources are
significant~\citep{TGJ01}. Ideally one would survey the CMB fields at
the same frequency as the CMB observations with sufficiently high
sensitivity and resolution to detect all the sources that would
contaminate the power spectrum. This would require a telescope rather
more expensive than the VSA itself. Instead, we adopt a two-part
strategy based on existing facilities; VSA fields are surveyed at
15~GHz with the Ryle Telescope (RT)~\citep{waldram-02}, and then each of
the sources found is monitored at 34 GHz by a source subtractor
interferometer located next to the VSA. More details of the source
subtraction procedure are given in Paper II.

The source subtraction interferometer consists of two 3.7-m diameter
dishes fitted with the same receivers as the main array, and equipped
with identical IF and correlator. The dishes are mounted such that they
track with the same rotation of parallactic angle as the main array,
so that polarized sources will be observed in the same
orientation with respect to the polarization of the feeds. The
antennas are housed in separate ground screens identical to the main
array ground screen, to limit ground pickup and to provide shielding
from the wind. The baseline is 9 m, giving a resolution of 4 arcmin in
a 9 arcmin field, which will not resolve any of the sources we
observe, while completely resolving out the CMB.

\subsection{The site}

The VSA is located at the Teide Observatory, Tenerife at an altitude
of 2400~m. The observatory is well established and, although on-site
support is always available, we are able to control both the telescope
and observations remotely.  The site provides excellent observing
conditions and has been well-tested for atmospheric seeing over a
15-year period. The inversion layer lies below the observatory for 75
per cent of the year; the twice-daily meteorological balloon flights
show that the precipitable water vapour is as low as 2 mm for 30 per
cent of the launches. For observations in the region $26-36$ GHz, the
typical transparency is 95 per cent. 
Correlated signals due to atmospheric emission
decrease with higher angular resolution. For example, at 33
GHz, the $8^{\circ}$ Tenerife beam-switch experiment (for example,
\citet{davies-96}) was affected by water vapour fluctuations for 80
per cent of the time while the Jodrell Bank-IAC 33 GHz
interferometer~\citep{melhuish1999} was affected for 20 per cent of the time at
$\sim 2^{\circ}$ resolution ($\ell \sim 110$) and $\le10$ percent at
$\sim 1^{\circ}$ resolution ($\ell \sim 210$). On the VSA baselines
($\ell\approx 150-900$) we expect even less loss of data due to atmospheric
effects. In fact, during our first season of observations (September
2000 -- September 2001), the site proved to have exceptional
conditions; only 5 percent of the data were rejected due to weather.

\subsection{Comparison with other CMB instruments}

An important factor in assessing the confidence we can place in measurements
as difficult, yet as important, as the CMB power spectrum, is the agreement
between experiments with different methods that are susceptible to different
systematic effects. The VSA is similar in its sensitivity and angular coverage
to several other CMB experiments that have reported results recently;
BOOMERanG \citep{netterfield-01}, MAXIMA \citep{hanany-2000}, DASI
\citep{halverson-02} and 
CBI~\citep{cbi}. Here we consider briefly the important systematic
effects in the 
VSA in comparison to these experiments.

BOOMERanG and MAXIMA are focal-plane arrays of bolometers on small, off-axis
telescopes suspended from stratospheric balloons. They have very high
instantaneous sensitivity at frequencies where the total galactic and
extragalactic foregrounds are at their minima, and they operate above most of
the atmosphere, eliminating that as a source of contaminating emission. Their
main problems are concerned with calibration; the uncertainties in the shapes
and areas of the beams from each detector, and their pointing on the sky,
limit the ability to recover the power spectrum at resolutions close to the
beam size. Recovery of the power spectrum at very low $\ell$ is
limited by receiver 
stability during each scan of the telescope.
There is also an overall uncertainty in the temperature scale due to
the lack of knowledge of the beam areas and, particularly for BOOMERanG, the
lack of good flux calibration sources observed during the flight. The quoted
uncertainty for the BOOMERanG power spectrum due to the temperature
calibration is 20 percent, in addition to the beam uncertainty
\citep{netterfield-01}. Foreground contamination at 150 GHz is small, and
dominated by galactic dust emission, the effect of which declines at higher
$\ell$. 

Calibration of instruments such as the VSA is much less problematic. Flux
calibrators with well-studied properties are available (such as Cas A or
Jupiter) and these can be observed repeatedly, and cross-checked with other
secondary calibrators. 
Overall calibration error is dominated by the uncertainty in the
absolute temperature of Jupiter, which is about 3.5
percent~\citep{mason_casscal}, leading
to a 7 percent error 
in the power spectrum. The VSA is much less sensitive than the balloon
experiments, however, with slightly worse random errors from 500 hours
observation than MAXIMA's from 3 hours. The main foreground contaminant is
extragalactic point sources, whose effect increases as $\ell^2$. The dominant
errors and contaminants are thus quite different from the balloon experiments,
so agreement between VSA, BOOMERanG and MAXIMA would be a very good sign of
the believability of the current power spectrum measurements.

Much the same arguments apply to the results from the other
interferometers, DASI and CBI, which are very similar to the VSA in
many respects, having similar numbers of antennas and the same
observing frequency. However, there are significant differences
between the three interferometer experiments which also allow for
independent cross-checks of calibration and systematics. The VSA and
CBI use the same primary flux calibrators; DASI, located at the south
pole, is unable to see these and instead uses an absolute measurement
of system temperature based on hot and cold loads. CBI, and DASI in
its first season, had no ground screen, and hence suffered from
significant pick-up of correlated emission from the ground, which had
to be removed by differencing fields observed at the same ground
angles. The VSA avoided this problem but suffered from a spurious
correlated signal related to cross-coupling between the closest-spaced
antennas, which is removed by Fourier filtering of the time-ordered
data (see section \ref{filtering}).

The biggest design difference between the CBI and DASI on one hand and
the VSA on the other is that the VSA has semi-independent tracking
rather than fully co-mounted antennas. The resulting variation of path
length to a source as the earth rotates imposes a modulation on the
signal which can be used to very effectively reject signals from
elsewhere in the sky, for example from the Sun or the Moon. This
allows the VSA to observe during daylight, improving the observing
efficiency. (Note that on a typical baseline there are 50 fringes
during an observation 
of a primordial field.)

All of the current CMB experiments thus have different calibration and
systematic problems, markedly so in the case of the interferometers
versus the balloons, and therefore agreement between them is essential
to establish the reliability of the power spectrum measurements.

\section{CALIBRATION AND DATA REDUCTION}

\subsection{Amplitude and phase calibration}

Absolute flux calibration of VSA observations is based on the flux scale of
\citet{mason_casscal}.  Our primary flux calibrator is Jupiter, for which the
brightness temperature is taken to be 154.5~K at our observing frequency of
34~GHz. The solid angle of the planet is calculated at the epoch of each
calibration observation and the flux determined. Jupiter is unresolved on all
VSA baselines (including the source subtractor baseline) so no further
correction is made for the angular size. We also observe Tau A and Cas A,
whose fluxes are determined via our primary calibration of Jupiter.
Transferring our calibration scale to both Cas A and Tau A ensures
that 
an alternative bright calibrator source is always available should a Jupiter
observation be lost due to, for example, bad weather, or obscured by the Sun
or Moon.  Care has to be taken, however, when transferring our flux
calibration to Tau A, since it is known to be polarized (approximately
8~percent 
at 34 GHz, \citep{polarization}), and hence the flux as observed by the VSA
changes with the hour angle of observation.  To ensure that this does not
introduce any systematic error, our calibration observations of Tau A are
short and are made at the same hour angle each day.

We also make daily observations of two fainter sources, 3C48
and NGC7027, allowing us to check the quality of observations
throughout the observing day.  The measured flux ratios of these
sources to our primary calibrators agree well with those reported by
Mason et al., suggesting that the accuracy of our flux calibration is
dominated by the error estimate of the absolute temperature of
Jupiter, which is approximately 3.5 percent.  Further to this, we have
cross-checked our calibration with that used by the CBI instrument
\citep{cbi}.  CBI also uses Jupiter as one of its primary calibrators
and, since they have a bandwidth of 10 GHz centred on 31~GHz, this has
enabled them to estimate the spectral index of Jupiter and Saturn over
the frequency range 26-36~GHz and hence cross-check our calibration at
34~GHz. We have cross-calibrations based on Jupiter and Saturn which
agree to 1 percent.

Phase calibration of the VSA is also applied on a daily basis using the same
three calibrators, Cas A, Tau A and Jupiter. The telescope is sufficiently
phase stable that a single phase calibration made within $\pm 12$ hours of a
CMB observation is adequate to calibrate the phases to $<10^{\circ}$. In the
same process, errors in the quadrature of up to $15^{\circ}$
between the real and imaginary parts of
each visibility are corrected for. 
These quadrature errors arise due to path differences in the
cables going to the correlator boards.
Consider two measured components of a
visibility $\mathcal{R}$ and $\mathcal{I}$ at angles $\phi$ and $\psi$
respectively from the theoretical fringe-rotated components $\mathcal{R}''$
and $\mathcal{I}''$. The quadrature error is $\chi = \psi-\phi$. Components
$\mathcal{R}'$ and $\mathcal{I}'$ which are orthogonal to each other can be
synthesised by rotating by $\beta=\chi/2$ in opposite directions.  These can
then be rotated by $\alpha=\phi+\beta$ to give visibilities corresponding to
the appropriate phase centre. Thus:
\begin{gather*}
\mathcal{R}' = \frac{\mathcal{R}\cos\beta}{\cos{\chi}} + 
\frac{\mathcal{I} \sin{\beta}}{\cos{\chi}}\\
\mathcal{I}' = \frac{\mathcal{I}\cos\beta}{\cos{\chi}} + 
\frac{\mathcal{R} \sin{\beta}}{\cos{\chi}}\\
\mathcal{R}'' = \mathcal{R}'\cos{\alpha}-\mathcal{I}'\sin{\alpha}\\
\mathcal{I}'' = \mathcal{I}'\cos{\alpha}+\mathcal{R}'\sin{\alpha}\\
\end{gather*}

\begin{figure}
\epsfig{file=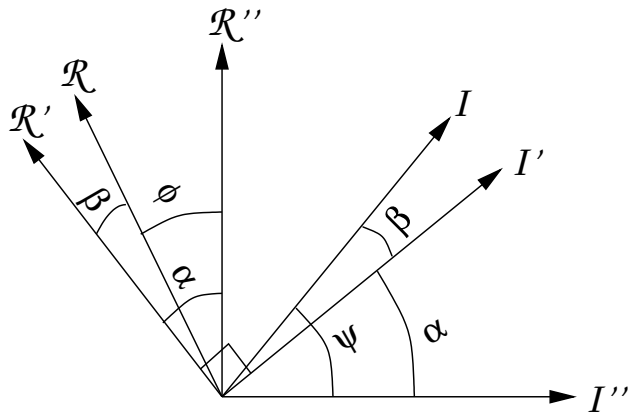,width=8.5cm}
\caption{\label{quadrature}Fringe rotation of visibilities with
quadrature errors.}
\end{figure}

Figure~\ref{quadrature} shows this scheme. Correcting for quadrature errors
necessarily results in a loss of signal to noise by a factor $\cos{\chi}$ and
the data must be down-weighted appropriately.

\begin{figure}
\epsfig{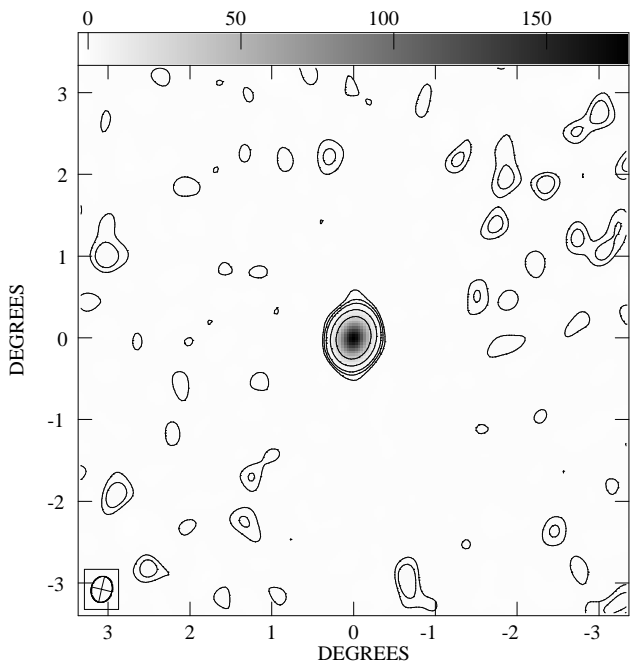}
\caption{An calibration observation of Jupiter, showing the dynamic range
possible in VSA maps. The contours are at 0.25 percent, 0.5 percent, 1 percent, 4 percent and 16 percent of
the peak flux of 176~Jy. The half-power synthesised beam is shown at the lower
left. The noise level is consistent with the expected thermal noise.}
\label{fig:x-calib}
\end{figure}

Figure \ref{fig:x-calib} shows a typical CLEANed map of a calibrator, in this case
Jupiter. Such maps are made daily to check the performance and stability of
the telescope; the amplitude and phase stability are such that any un-CLEANed
artefacts are below the thermal noise level. Since this observation has a
higher signal-to-noise ratio than any CMB measurement, we can be confident
that our CMB measurements are unaffected by dynamic range limitations.

\subsection{Pointing and geometry accuracy}

The pointing accuracy of the VSA is primarily determined by mechanical
alignment tolerances, but we frequently make long observations of unresolved
calibrators in order to check both the pointing and geometry of the array.
Pointing is checked with two sets of observations. Firstly, the antennas in
turn are offset in the antenna tracking co-ordinate on either side of the
calibration source; secondly, the whole table is offset in the table tracking
co-ordinate. These allow us to fit for offsets in the table and antenna
encoders, and to check the co-alignment of the antennas. Alignment is better
than 5 arcmin, about 1.5  percent of the beamwidth.

Geometry is checked using long observations of two or more calibrators at
different declinations. In conjunction with a model of the telescope, we
employ a maximum-likelihood technique to fit simultaneously for $\sim$400
parameters.  These include the $x$, $y$ and $z$ co-ordinate of each antenna,
complex gains for each of the 182 correlator channels and the effective
observing bandwidth and frequency of each baseline. These parameters can be
determined accurately enough to ensure that in subsequent CMB observations,
phase and amplitude errors due to geometry errors are always less than the
thermal noise.

\subsection{System temperature calibration }

The overall gain of each antenna is monitored via a noise injection
system. A modulated noise signal is injected into each antenna via a
probe in the horn, and is later measured using phase-sensitive
detection after the automatic gain control stage in the IF system.
The relative contribution of the constant noise source to the total
output power from each antenna varies inversely with system
temperature, and thus a correction can be made to the overall flux
calibration.  
Assuming that the receiver temperature is constant, this system allows
us to account for variations in atmospheric attenuation of the
astronomical signal. 
It provides an excellent indication of the
weather conditions and is used as a primary indicator for flagging
data. For good observing conditions, the gain corrections applied
using this system are typically less than a few percent.

\subsection{Calibration of the source subtractor}

Calibration of the source subtraction baseline is similar to that of the VSA
main array.  The primary flux calibrator for the source subtractor is NGC7027,
whose flux is taken to be 5.45 Jy at 34~GHz.  Observations of NGC7027 are made
each day, in addition to observations of 3C48, 3C273 and Jupiter.  The much
higher flux sensitivity of the source subtractor makes it difficult to observe
the same bright calibrators as the main array without using the low-gain
setting of the correlator.  Cross-calibration of the VSA and source subtractor
is therefore achieved by low-gain observations of Jupiter; the gain switch is
calibrated by observations of 3C273 in both gain states.  Agreement between
the VSA and source suntractor is found to be better than 3 percent.

Phase calibration of the source subtractor differs slightly from that for the
main array.  Since the source subtractor continually slews between pointed
observations of point sources in the VSA fields, it is important to check the
phase and pointing stability of source observation throughout an observing
run.  For this reason, between each set of 10 pointed source observations, we
observe a phase calibrator source.  The sources chosen to be phase calibrators
need only be bright enough to get good signal-to-noise ratios 
in a 400 second
integration, and are typically chosen to be the brightest sources already
identified within a VSA field.

As with the main array, the long-term pointing accuracy and stability of the
source subtractor baseline is monitored via long calibrator runs utilising the
full range of hour angles.

\subsection{Data reduction}

The correlated signal from each of the 91 VSA baselines is sampled each second
and, before the raw data are calibrated and smoothed, a series of preliminary
flagging operations are performed.  These are based on the status of the
telescope each day, for which our main diagnostic tools are the system
temperature monitor and automatic gain control levels, both of which monitor
the power level from each antenna during an observation.  This information can
be used to flag for bad weather or warm cryostats. We also independently
monitor the temperature of each cryostat, but the system temperature can be
more sensitive and is useful as an early warning of problems with the
cryogenics. Pointing errors are recorded throughout each observation, and the
data are automatically flagged if the pointing error is greater than
0.5$^\circ$.  Missing or repeated data samples, due to
communication glitches between the correlator readout micro and the data
recording computer, are also corrected. Finally, times at which the field
centre is within 40$^\circ$ of the Sun or within 30$^\circ$ of the Moon are
flagged, and a pre-analysis cut on the data at a 10-$\sigma$ level is
applied. Typically, up to 3 percent of the data on each field is flagged in
this manner.

As well as the fast ($\sim 1$~kHz) phase switching that is demodulated in
hardware in the correlator, we also employ a slow phase switching sequence
with a cycle time of 16s that is demodulated in software and removes any
offsets in the hardware integration and sampling. After the initial flagging,
the data are demodulated and smoothed over 16 seconds.  Since this smoothing
occurs before the data have been fringe rotated, a small amplitude correction
must be applied to compensate for the reduction in amplitude of the raw
fringes.  The data can then be filtered for contaminating signals.  As
described in Section \ref{filtering}, the effects of contaminating signals,
weather and bright sources outside the beam, can be effectively removed via
the application of a high pass Fourier filter.  For a typical VSA field
observation, up to 20 percent of the data can be lost due to filtering, the
majority of which is on baselines shorter than 30~$\lambda$.  After filtering,
the data are fringe-rotated to the centre of the field and
calibrated. Phase calibration and quadrature correction 
are also implemented at this point. 
The system temperature correction
is applied after the primary calibration, correcting for variations in the
gain of each antenna and for changes in the atmospheric attenuation. Finally,
the data are averaged to 64-second samples and a post-analysis chop at the
5-$\sigma$ level ($\sim 150$~Jy) is applied.

Although each day's data must be examined and flagged individually, the
basic data-reduction process outlined above can be applied
automatically to the data for each field.  As a further check on the
data analysis, maps are made of both the calibrator and field
observations for each day.  The noise level on each map is checked to
be consistent with that expected, and errors in the
data analysis are easily detected.

\section{CONTAMINANTS}

As well as the CMB signal and thermal noise, the VSA data also include
several types of contaminating signals, including bright sources such
as the Sun and Moon, as well as instrumental artefacts. Here we
describe the techniques for eliminating these from the data and
checking the data for residual contaminants.

\subsection{Spurious signal}

\begin{figure}
\epsfig{figure=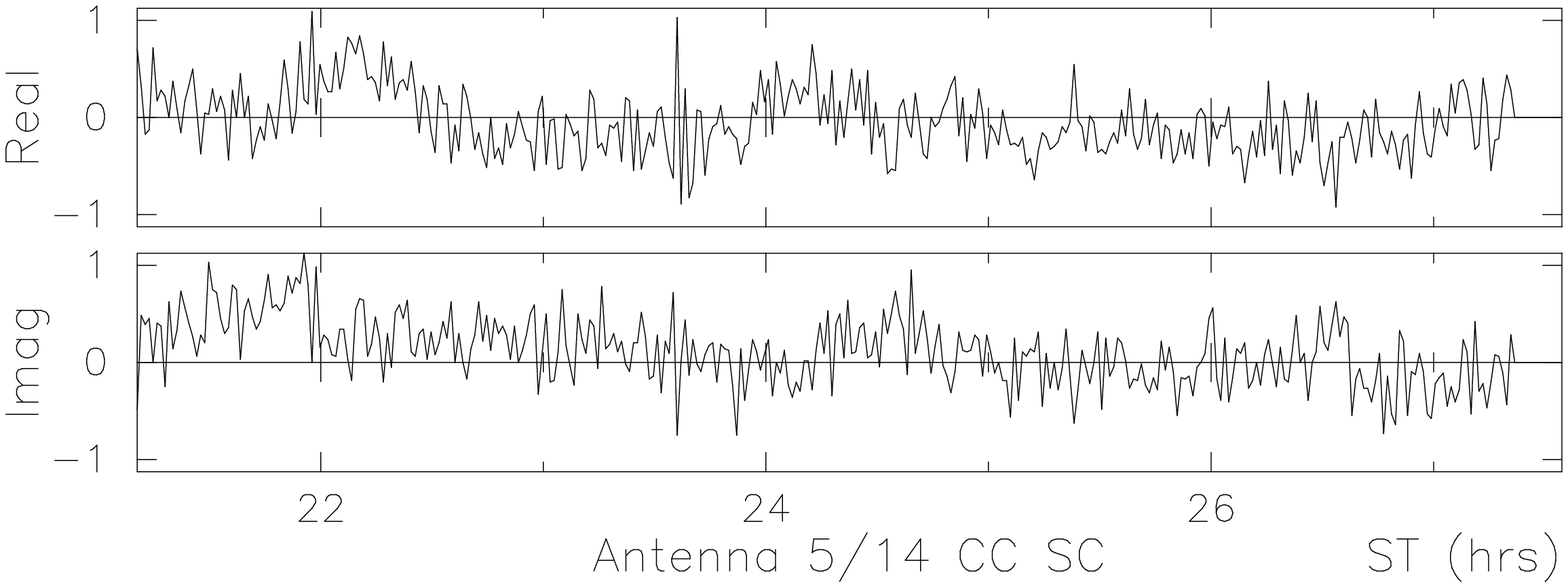,width=8.5cm}
\epsfig{figure=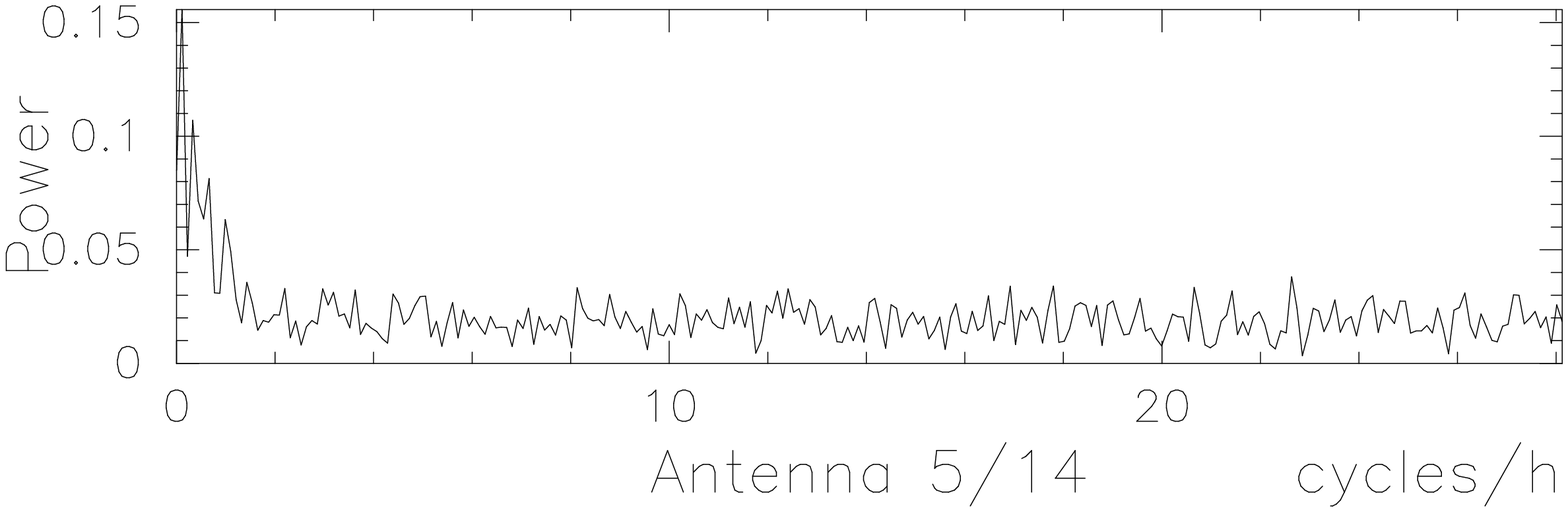,width=8.5cm}
\caption{Typical spurious signal data on a short VSA baseline for an
observation of a blank patch of sky, shown
both as complex time series data and as a power spectrum. \label{spur}}
\end{figure}

Commissioning observations of blank fields showed that the VSA was detecting a
correlated signal which did not originate from the sky (see Figure~\ref{spur}
for an example of this signal and its power spectrum).

Further investigation determined the following characteristics for the
signal:

\begin{enumerate}

\item It is reasonably
constant if the telescope is held stationary.
\item It is broad-band.
\item Its strength varies dramatically with baseline, peaking at
approximately 1~mK on the shortest baselines but being undetectable on
most longer baselines.
\item The signal vanishes if a large aluminium sheet is inserted
between the antennas, indicating that the signal enters through the
antennas and is related to coupling between them.
\item The signal does not vary when the tilt-table is driven in
elevation, showing that the signal source is on the table rather than,
for example, being radiation diffracting over the ground screen.
\item If the antennas are driven, the signal is modulated at a rate consistent
with the rate of change of path between adjacent edges of the
antennas.  This means that during an observation of an astronomical
source the maximum possible rate of change for the spurious signal is
about 2 cycles per hour.

\end{enumerate}

This spurious signal is not consistent with radiation from the receivers being
coupled between the antennas, as the measured coupling even at the shortest
spacings is less than $-100$~dB, which would imply a correlated noise
temperature of several hundred Kelvin being emitted by the receivers. 
Comparable levels of coupling have been measured by \citet{padin-00}
for the CBI telescope. 
Possible explanations for the source include thermal radiation from the antenna
surface, or scattering of ambient radiation off one antenna that couples to
another, an effect seen in the Australia Telescope Compact Array
(R. Subrahmanyan, priv comm).  
However,
the signal can be effectively removed from the data by suitable Fourier
filtering of the time-ordered data.

\subsection{Filtering according to astronomical fringe rate} \label{filtering}
\label{sec:filtering}
\begin{figure}
\psfig{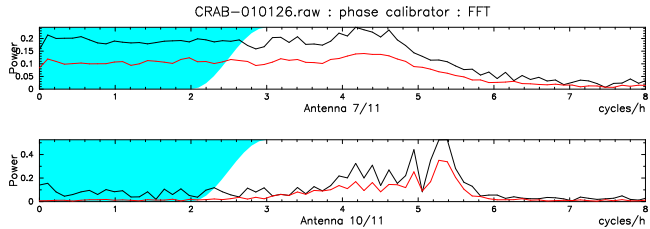}
\caption{Power spectra from VSA observations of a bright point source observed
by the VSA with two different baselines. In each plot the theoretical power
spectrum is shown as a light line, the measured power spectrum as a heavy
line. The shaded region represents the Hanning filter which will remove any
low frequency spurious signal from the data (in the upper plot the spurious
signal is swamped by the source). Data which corresponds to fringe rates less
than 3 cycles per hour must be flagged since it lies within the range of the
filter. In the upper plot this means that some data are lost; in the lower
plot no data are lost. \label{filt}}
\end{figure}

Because of the semi-independent tracking of the VSA antennas, the correlated
astronomical signal is modulated as the relative path length to the antennas
changes. This modulation is typically much faster than the observed rate of
change of the spurious signal. We can therefore filter out the spurious signal
using a high-pass Fourier filter on the time-ordered data. We use a Hanning
(cosine-edged) filter in order to minimise ringing, ie long-period
correlations, in the time domain (Fig.~\ref{filt}).

\begin{figure}
\epsfig{file=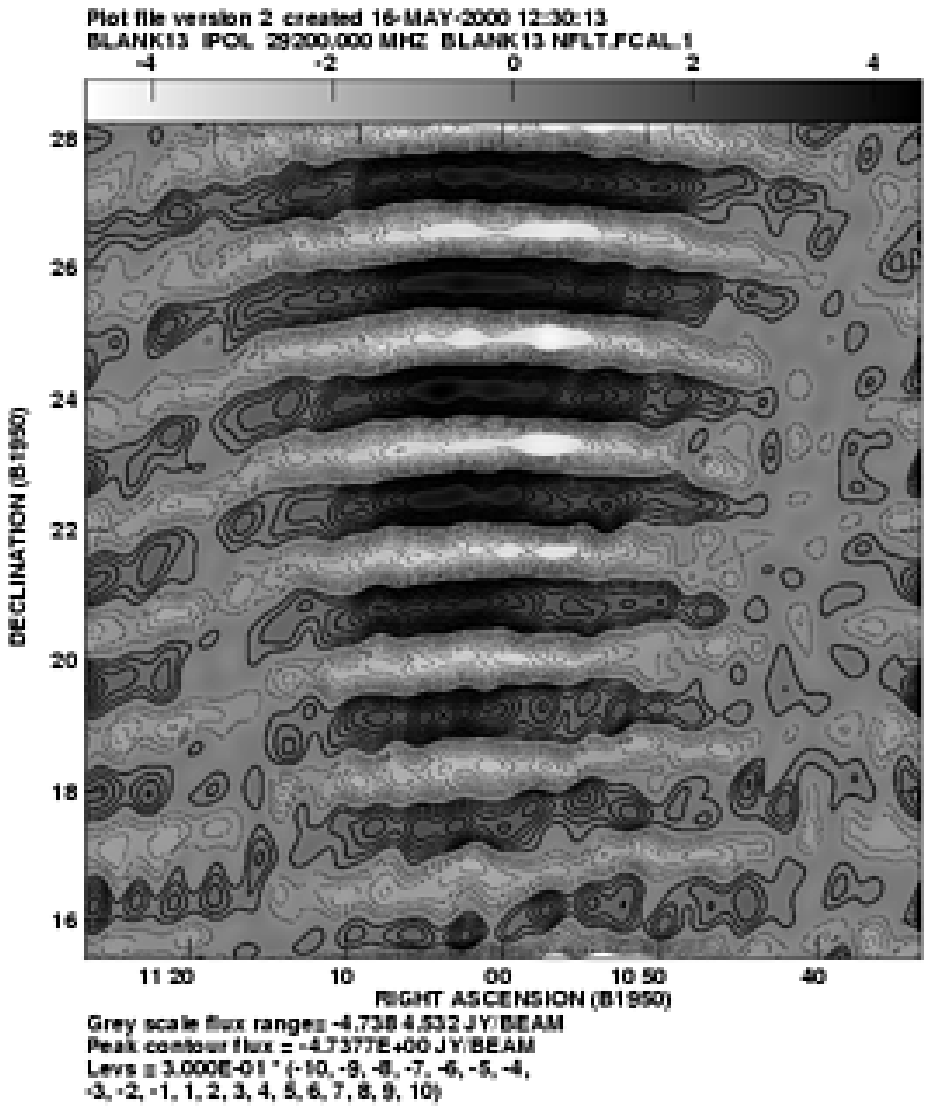, width=7cm, clip=}
\epsfig{file=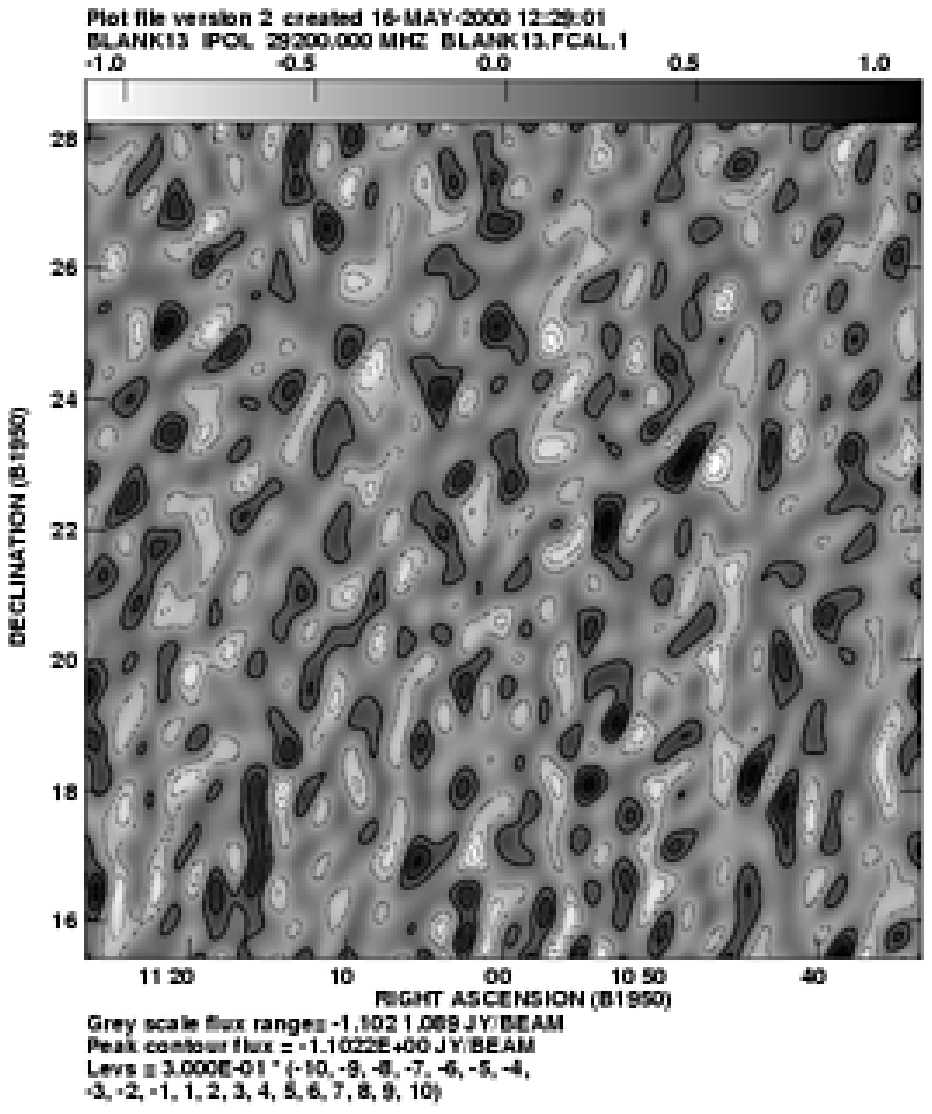, width=7cm,clip=}
\caption{Maps of a single 6 hour observation of a blank field with and
without filtering; the rms of the filtered map is consistent with the thermal
noise level of the telescope.\label{filt-nofilt}}
\end{figure}

Although most of the time the astronomical signals are modulated faster than
the filter frequency and are thus not affected by it, at some times the fringe
rate drops into the filter stop band; this depends on the baseline
orientation, source position and hour angle. We flag out any data taken when
the astronomical fringe rate lies within the cutoff range of the filter; in
general this leads to approximately 20 percent of the visibilities for an
observation being discarded. Figure \ref{filt-nofilt} shows a single
observation of a blank field with and without filtering; the filtered map is
consistent with the thermal noise level.

We have studied the effect of this filtering on the data analysis by
simulating CMB observations from sky realisations with $\Lambda$CDM power
spectra, following exactly the procedure used for real observations
all the way through to maps and recovered power spectra, and have
shown that any correlations introduced in the data by the filtering
have negligible effect on the results.

\subsection{Filtering of the Sun and Moon}

It is possible to use the technique described above to remove the
effects of very bright sources which lie well away from the field of
view but are still detected in the sidelobes of the telescope
beam. The source will in general have a different astronomical fringe
rate to the field being observed. We therefore fringe rotate the data
to the bright source, apply a high pass Fourier filter, fringe rotate
back to the field centre, and flag any data where the fringe rate of
the field lies within the range of the filter.

For the VSA the only sources that are bright enough to affect
observations when they fall in distant sidelobes are the Sun and the
Moon, which have approximate flux densities of 10~MJy and 0.3~MJy
respectively at 34~GHz. For our observations of the CMB we do not
observe any fields within $40^{\circ}$ of the Sun or $30^{\circ}$ of
the Moon, and apply the above filtering technique whenever these
sources are above the VSA ground screen. 

\begin{figure}
\epsfig{figure=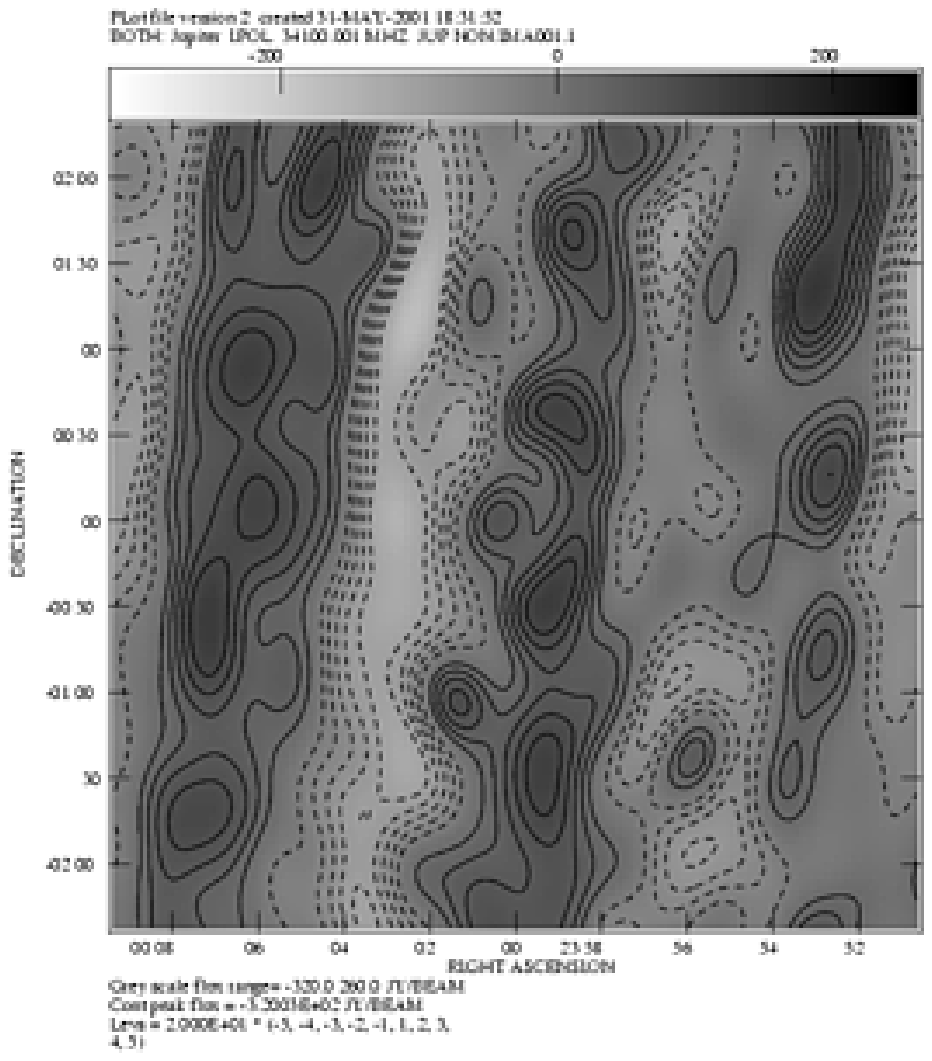,width=7cm,angle=0,clip=}
\epsfig{figure=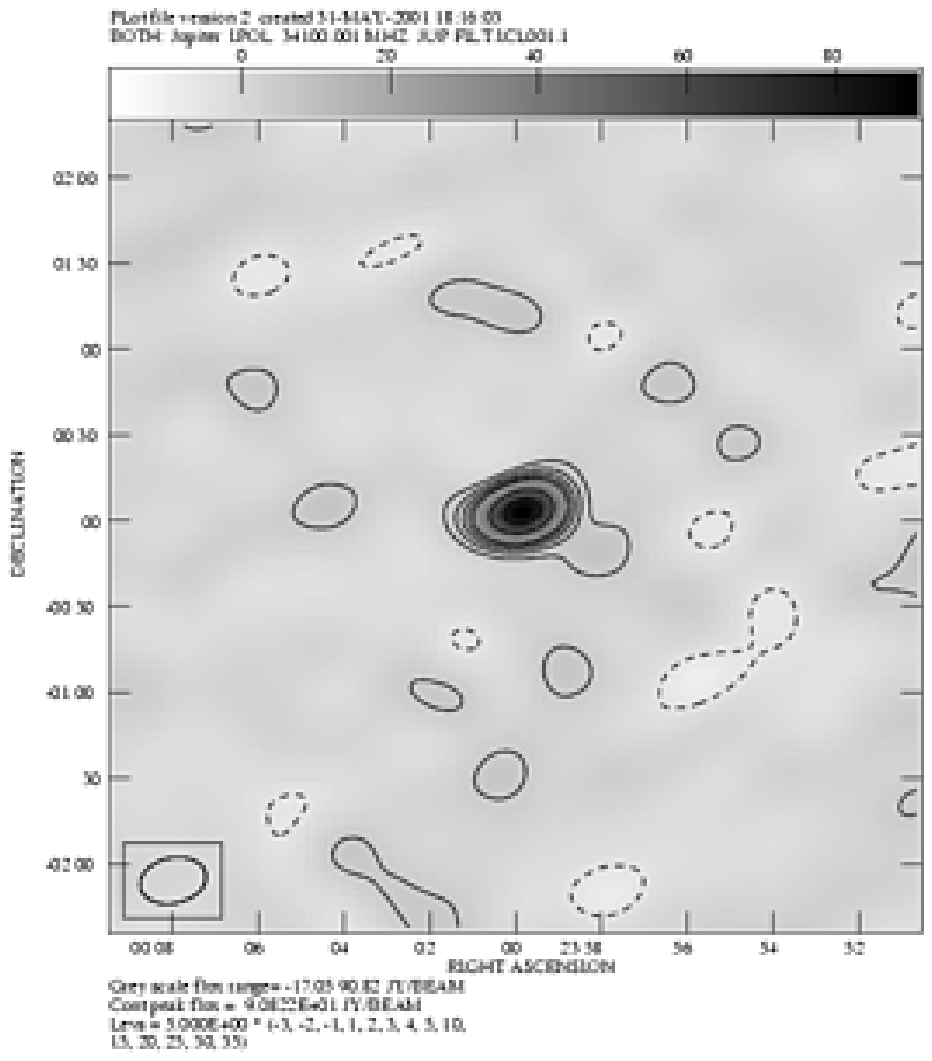,width=7cm,angle=0,clip=}
\caption{VSA observation of Jupiter (flux density $\approx90$ Jy) with the Sun
(flux density $\approx10$ MJy) at $11^{\circ}$ distance. Note that the
FWHM of the primary beam is
$4.6^{\circ}$. The two plots show the map before and after filtering,
with contour levels of 20~Jy and 5~Jy respectively.\label{jup}}
\end{figure}

\begin{figure}
\epsfig{file=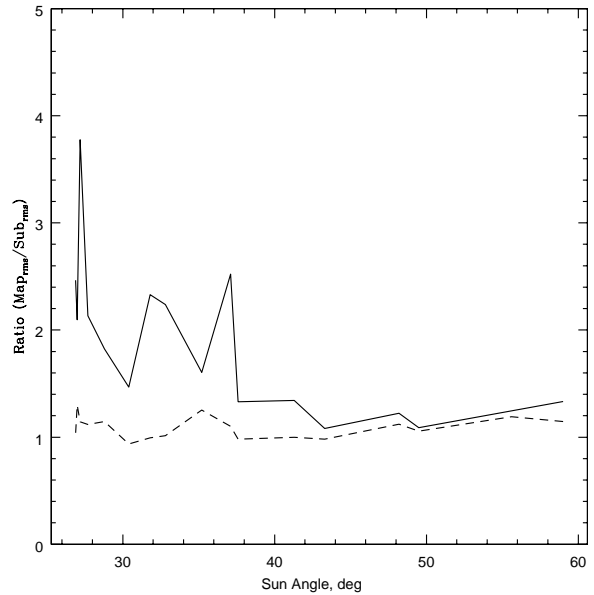, width=8.5cm}
\caption{Plot of the rms noise relative to the thermal noise versus
angle from the Sun for filtered (dashed line) and unfiltered 
(solid line) data.}
\label{sun}
\end{figure}

The power of the technique is
demonstrated in Figure \ref{jup} where Jupiter, with flux density of
approximately 90~Jy, is observed within $11^{\circ}$ of the Sun.
The filtered map shows that the contaminating effect of the Sun has been
successfully removed and that the only significant feature in the map is a
point source corresponding to Jupiter. Its flux density is as expected
and the rms on the map is
consistent with 
the thermal noise. We have examined the rms noise level on maps as a function
of distance from the Sun with and without filtering. Figure \ref{sun} shows
the rms noise levels per unit observation time on a sequence of 6-hour blank
field observations at various distances from the Sun, both with and without
filtering. Excess noise due to the Sun can be seen at angles less than
$40^{\circ}$ in the unfiltered case, but no excess noise is seen in the
filtered data (although of course more and more data has to be flagged the
closer the observation is to the Sun). For CMB observations, we do not use
data taken less then $40^{\circ}$ from the Sun, {\em and} we employ filtering.

\section{Final commissioning observations}

\begin{figure}
\epsfig{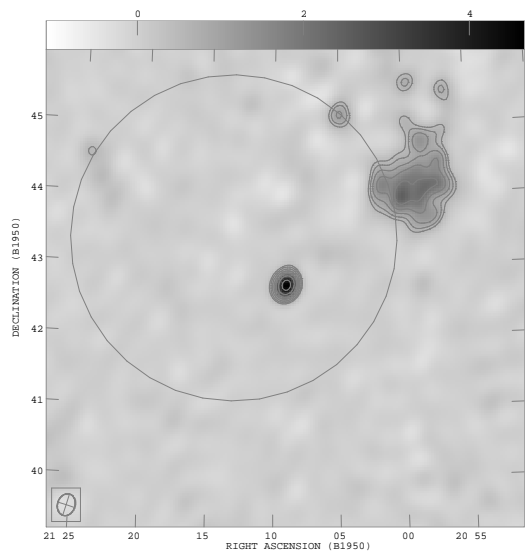}
\caption{Offset pointed observation of NGC7027. The circle shows the
half-power limit of the primary beam. The reduction in apparent flux of the
source is consistent with the calculated primary beam. Also visible in the
image is some diffuse galactic structure; this is taken into consideration
when using NGC7027 as a cross-calibrator between the VSA main array and the
source subtractor. The contour interval is 0.35 Jy~beam$^{-1}$\label{ngc}}
\end{figure}
As a final checks of the telescope function we observed two regions in the
Galactic plane. We observed the source NGC7027 with the telescope pointing and
phase centres offset from the source position. This tests all the fringe
rotation, calibration and mapping algorithms, as well as confirming the
expected attenuation due to the primary beam. Also visible in the image
(Figure \ref{ngc}) is some diffuse galactic emission (part of a larger
structure that is cut off by the primary beam attenuation). This emission lies
well outside of the primary beam of the source subtraction telescopes, which
use NGC7027 as their primary calibrator.

\begin{figure}
\epsfig{file=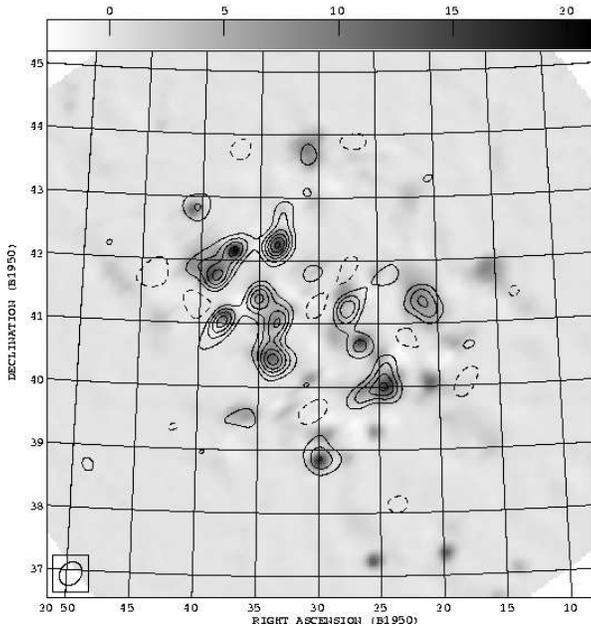,width=8.5cm} \caption{Commissioning observation of
  Cygnus-X. Contours (at 3~Jy) are VSA observations at 34~GHz. Note that
  the effect of the VSA primary beam (FWHM=$4.6^{\circ}$) has not been
  removed from this map. The 
  grey scale is data at 14.35~GHz from Green Bank \citep{cyg_loop}. }
\label{fig:cygnus}
\end{figure}

We also observed the Cygnus-X region, an area of bright 
free-free extended emission, for
which an independent observation with similar resolution at a close frequency
was available. Fig~\ref{fig:cygnus} shows our observed 34~GHz contours are
overlaid on 14.35~GHz Green Bank data~\citep{cyg_loop}. There is very good
agreement between the structure observed at the two frequencies. 

\section{CONCLUSIONS}

We have tested the operation of the VSA by observations of calibrators, blank
sky, and extended galactic structure. The telescope has been demonstrated to
work according to specification. In particular, we have shown that systematic
effects such as the effects of the Sun and Moon, and coupling between the
antennas, can be removed from the data to a very low level, allowing reliable
long integrations on CMB fields. We have also shown that the data can be
calibrated to an accuracy limited only by the absolute flux measurement of our
primary calibrator, and that our separate telescope for observing point radio
sources can be accurately cross-calibrated with the main array.

\section*{ACKNOWLEDGEMENTS}

We thank the staff of the Mullard Radio Astronomy Observatory, Jodrell Bank
Observatory and the Teide Observatory for invaluable assistance in the
commissioning and operation of the VSA. The VSA is supported by PPARC and the
IAC. Partial financial support was provided by Spanish Ministry of
Science and Technology project AYA2001-1657.
A. Taylor, R. Savage, B. Rusholme, C. Dickinson acknowledge
support by PPARC 
studentships. K. Cleary and J. A. Rubi\~no-Martin acknowledge  Marie
Curie Fellowships  of the European
Community programme EARASTARGAL, ``The Evolution of Stars and Galaxies'', under
contract HPMT-CT-2000-00132. K. Maisinger acknowledges support from an EU Marie
Curie Fellowship. A. Slosar acknowledges the support of
St. Johns College, Cambridge. We thank Professor Jasper Wall for
assistance and advice throughout the project.

\bibliography{cmb_refs}
\bibliographystyle{mn2e}
\bsp 

\label{lastpage}

\end{document}